\documentclass[useAMS,usenatbib]{mn2e}

\title[Properties of semi-convection and convective overshooting]
{Properties of semi-convection and convective overshooting for massive stars}
\author[Ding \& Li]
       {C.Y. Ding$^{1,2,3}$ and
        Y. Li$^{1,2}$
\\
$^1$Yunnan Astronomical Observatory,Chinese Academy of Sciences,Kunming \\ 650011,China;dcy84310@mail.ynao.ac.cn,ly@ynao.ac.cn\\
$^2$Key Laboratory for the Structure and Evolution of Celestial Objects,Chinese Academy of Sciences,\\Kunming 650011,China\\
$^3$Graduate University of Chinese Academy of Sciences, Beijing 100049, China\\
}

\begin{document}

\date{}

\pagerange{\pageref{firstpage}--\pageref{lastpage}}
\pubyear{2012}

\maketitle

\label{firstpage}

\begin{abstract}
Properties of semi-convection and core convective overshooting of stars with 15 $M_{\odot}$ and 30 $M_{\odot}$ are calculated in the present paper. New methods are used to deal with semi-convection. Different entropy gradient is used when adopting the Schwarzschild method and the Ledoux method which are used to confine the convective boundary and to calculate the turbulent quantities: $\frac{\partial \overline{s}}{\partial r}=-\frac{c_{p}}{H_P}(\nabla-\nabla_{\rm ad})$ when the Schwarzschild method is adopted and $\frac{\partial \overline{s}}{\partial r}=-\frac{c_{p}}{H_P}(\nabla-\nabla_{\rm ad}-\nabla_{\mu})$ when the Ledoux method is adopted.      Core convective overshooting and semi-convection are treated as a whole part and the development of them are found to present nearly opposite tendency, more intensive core convective overshooting lead to weaker semi-convection. The influences of different parameters and the convection processing methods on the turbulent quantities are analyzed in this paper. Increasing the mixing-length parameter $\alpha$ leads to more turbulent dynamic energy in the convective core and prolonging the overshooting distance but depressing the development of semi-convection. The Ledoux method adopted leads to overshooting extending further and semi-convection developing suppressed.
\end{abstract}


\section{Introduction}

\textbf{Convection is commonly present in stars. The Schwarzschild criterion is valid in a chemically homogeneous region. When a chemical gradient is present, convection can only develop if the Ledoux criterion is violated.} Convective overshooting and semi-convection \citep{b30} significantly affect the structure and evolution
of massive stars. The overshooting beyond the convective core extends the mixing range of chemical
elements, which supplies more nuclear fuel to the central hydrogen burning. On the other hand, however,
the semi-convection affects the efficiency of chemical mixing beyond the convective core, resulting
 in important effect on the future shell hydrogen burning.

According to the Schwarzschild criterion, the boundary of the convective core is fixed at where the
radiative temperature gradient is equal to the adiabatic temperature gradient. However, a convective
cell may have a velocity unequal to zero when it arrives at the edge of the convective core, and will
still go further into stably stratified region. Such a phenomenon is referred to as the convective
 overshooting. Outside the convective core, the turbulent velocity decelerates quickly to zero.
 This region is named as the overshooting region. The overshooting beyond the convective core results
 in significant effects on the main sequence evolution of massive stars \citep{b6, b7, b8, b9, b3, b10, b11, b24}. Moreover, the overshooting can also happen outside convective
 shells, but its effect has not been studied systematically. The convective overshooting should be
  apparently treated by nonlocal convection theories \citep{b32, b4, b14, b5}. However, different nonlocal convection models usually result in different overshooting
  lengths. In many applications, the overshooting beyond the convective core is implemented in a
  parameterized way: a fixed overshooting length being used, and its value being adjusted in order for
  the resulted stellar models to be in agreement with observations.

In massive stars, the stratification outside the convective core is sometimes unstable according
to the Schwarzschild criterion but stable according to the Ledoux criterion. Such a region is the
so-called semi-convection zone. \citet{b17} pointed out that the semi-convection zone is pulsational
unstable, and suggested that the motion carries no heat but may result in incomplete mixing of the
chemical elements on the thermal time-scale. Theoretical models \citep{b28} and 2D numerical
simulations \citep{b22} result in quite different results for the mixing efficiency.
The most popular method to treat such a mixing in the semi-convection zone is the diffusion
model \citep{b19}, which tends to recover the neutrality condition in the semi-convection
 zones.

The convective mixing \citep{b29}, both in the overshooting regions and in the semi-convection zones,
can be approximated by a diffusion process with appropriate estimation of the diffusion
coefficient \citep{b12, b31}. Due to large scales of the motions
and low viscosities of the stellar matter, convective motions in stars most probably run
 into turbulence. As a result, turbulent diffusivity dominates the convective mixing process.
 It is therefore desirable to investigate the turbulent properties outside the convective cores
 of the massive stars. For example, \citet{b20} studied recently the properties of turbulence
 below the bases of the convective envelopes for RGB/AGB stars by use of a turbulent convection
  model (TCM) proposed by \citet{b21}.

In this paper, we apply a modified version of Li \& Yang's TCM to the convective cores of the
massive stars, in order to investigate the turbulent properties in the overshooting regions
and semi-convective zones. In Sect. 2, we introduce basic equations of the TCM we have used.
 Model parameters and the input physics in our stellar models are put in Sect. 3. Then we
  discuss the obtained results for a $15M_{\odot}$ model in Sect. 4, and for a $30M_{\odot}$ model
   in Sect. 5. We summarize our main conclusions in Sect. 6.

\section{Turbulence models for convective overshooting and semi-convection}

The equations of the TCM we have solved are as follows:
\begin{eqnarray}
&&\frac{1}{\varrho r^{2}}\frac{\partial}{\partial r}(C_{s}\varrho r^{2}\frac{k}{\varepsilon}\overline{u_{r}'u_{r}'}\frac{\partial k}{\partial r})\nonumber\\
&&=\varepsilon+\frac{\beta g_{r}}{T}\overline{u_{r}'T'}\\
&&\frac{1}{\varrho r^{2}}\frac{\partial}{\partial r}(C_{s}\varrho r^{2}\frac{k}{\varepsilon}\overline{u_{r}'u_{r}'}\frac{\partial \overline{u_{r}'u_{r}'}}{\partial r})\nonumber\\
&&=\frac{2}{3}\varepsilon+\frac{2\beta g_{r}}{T}\overline{u_{r}'T'}+C_{k} \frac{\varepsilon}{k}(\overline{u_{r}'u_{r}'}-\frac{2}{3}k)\\
&&\frac{2}{\varrho r^{2}}\frac{\partial}{\partial r}(C_{t1}\varrho r^{2}\frac{k}{\varepsilon}\overline{u_{r}'u_{r}'}\frac{\partial \overline{u_{r}'T'}}{\partial r})\nonumber\\
&&=\frac{T}{c_{p}}\frac{\partial \overline{s}}{\partial r} \overline{u_{r}'u_{r}'}+C_{T}(\frac{\varepsilon}{k}+\frac{\lambda}{\rho c_{p}}\frac{\varepsilon^{2}}{k^{3}})\overline{u_{r}'T'}+\frac{\beta g_{r}}{T}\overline{T'^{2}}\\
&&\frac{1}{\varrho r^{2}}\frac{\partial}{\partial r}(C_{e1}\varrho r^{2}\frac{k}{\varepsilon}\overline{u_{r}'u_{r}'}\frac{\partial \overline{T'^{2}}}{\partial r})\nonumber\\
&&=\frac{2T}{c_{p}}\frac{\partial \overline{s}}{\partial r}\overline{u_{r}'T'}+2C_{e}(\frac{\varepsilon}{k}+\frac{\lambda}{\rho c_{p}}\frac{\varepsilon^{2}}{k^{3}})\overline{T'^{2}}
\end{eqnarray}

The dissipation rate $\varepsilon$ of the turbulent kinetic energy $k$ is modelled by:
\begin{eqnarray}
&&\varepsilon=\frac{k^{3/2}}{l}.
\end{eqnarray}
For the typical length of turbulence $l$, a model similar to the mixing-length theory (MLT) is adopted:
\begin{eqnarray}
&&l=\alpha H_{P}
\end{eqnarray}
where $\alpha$ is the so-called mixing-length parameter and $H_P$ the local pressure scale height. Furthermore, the total heat flux $F$ is composed of the radiative one $F_R$ and the convective one $F_C$:
\begin{eqnarray}
F=F_R+F_C=\frac{T\lambda}{H_P}\nabla +\rho c_P\overline{u'_r T'}
=\frac{T\lambda}{H_P}\nabla_{\rm rad},
\end{eqnarray}
where $\nabla$ is the actual temperature gradient and $\nabla_{\rm rad}$ the radiative temperature gradient. For more details please refer to \citet{b21}.

According to the MLT, the convection in the interior of massive stars can simply be handled as follows. Firstly, the convective boundaries are found by applying either Schwarzschild or Ledoux criterion. Secondly the temperature gradient is made equal to the adiabatic one, and thirdly the chemical elements are mixed homogeneously in the convection zones. Results are therefore different due to different criterion of convection to be used. According to the Schwarzschild criterion, the condition
\begin{eqnarray}
&&\nabla_{\rm rad}>\nabla_{\rm ad}
\end{eqnarray}
is used to determine the convection zones, where $\nabla_{\rm ad}$ is the adiabatic temperature gradient. According to the Ledoux criterion, however, the condition
\begin{eqnarray}
&&\nabla_{\rm rad}>\nabla_{\rm ad}+\nabla_{\mu}
\end{eqnarray}
is used to determine the convection zones, where
\begin{eqnarray}
&&\nabla_{\mu}=-(\frac{\partial \ln \rho}{\partial \ln T})_{\mu,p}^{-1}
                (\frac{\partial \ln \rho}{\partial \ln \mu})_{T,p}
                 \frac{d\ln \mu}{d\ln P}.
\end{eqnarray}
Moreover, the convective overshooting and the semi-convection are usually
treated as two independent processes. In most of massive stars' models, the
overshooting beyond the convective core is often adopted together with the Schwarzschild
 criterion; while the semi-convection is commonly used when the Ledoux criterion is applied.
 However, the problem of either convective overshooting and semi-convection has not been basically
 solved for the massive stars' evolution \citep{b18}.

In contrast with the MLT, the TCM's equations are based on the full equations of fluid hydrodynamics
and have to be applied to the whole stellar interior. As a result, there is no need to determine
the boundaries of the convective core and of the semi-convection zones. Instead, the averaged entropy
 gradient has to be used not only in the convective core but also in the overshooting regions and
 semi-convection zones. Particularly, in order to apply the TCM's equations in the overshooting regions
 and semi-convection zones, we consider two different methods to treat the averaged entropy gradient.
  The first method is:
\begin{eqnarray}
&&\frac{\partial \overline{s}}{\partial r}=-\frac{c_{p}}{H_P}(\nabla-\nabla_{\rm ad}-\nabla_{\mu}),
\end{eqnarray}
which will be referred to as the Ledoux method. It can be noticed that the chemical
gradient $\nabla_{\mu}$ is included in Eq.\,(11), which is in accordance with the Ledoux criterion.
Inside the convective core, the element abundances are homogeneous and the chemical
gradient $\nabla_{\mu}$ is zero. But outside the convective core, the mean molecular weight
decreases outward, and the chemical gradient contributes to increasing the entropy gradient.

It can be seen, by letting the left hand side of Eq.\,(3) be zero and inputting the
resulted $\overline{u'_r T'}$ into Eq.\,(1), that the averaged entropy gradient contributes
 respectively to either generation or dissipation of the kinetic energy of turbulence depending
 on its value being negative or positive, while the auto-correlation for the temperature fluctuation
 always contributes to the generation of turbulent kinetic energy. In the overshooting region just
 outside the convective core, the radiative temperature gradient is less than the adiabatic temperature
 gradient, which results in the averaged entropy increasing outward. Therefore, the chemical gradient
 further accelerates the increase of the averaged entropy gradient. On the other hand, however,
 the radiative temperature gradient is larger than the adiabatic one in the semi-convection zone even
 farther outside the convective boundary, but the chemical gradient can still keep the averaged
 entropy gradient positive. If the chemical gradient is no longer greater than the difference of
 radiative and adiabatic temperature gradients, the averaged entropy gradient will be negative
 and a full convection shell will be developed. It should be noticed that the averaged entropy
 gradient not only determines the boundaries of the convective core and convective shells, but also
 contributes to determine the properties of turbulence in the overshooting regions and semi-convection
 zones.

The second method is:
\begin{eqnarray}
&&\frac{\partial \overline{s}}{\partial r}=-\frac{c_{p}}{H_P}(\nabla-\nabla_{\rm ad}),
\end{eqnarray}
and it is referred to hereafter as the Schwarzschild method. This method is similar with those
commonly used approaches by adopting Schwarzschild criterion to determine the boundary of the
convective core and supplemented with the convective overshooting beyond the convective core.
\textbf{ The Schwarzschild method are used in $\mu$-gradient zone aiming to compare with the Ledoux method.}
Comparing results from these two methods, we can show more clearly the effects of the chemical
gradient on the properties of the overshooting beyond the convective core and of the semi-convection
zones.

\section{Model parameters and input physics}

All of our stellar models were computed by a stellar evolution code h04.f, which was originally
developed by Paczy\'{n}ski \& Kozlowski and updated by \citet{b27}. Nuclear reaction rates
are adopted from \citet{b2} and \citet{b15}. The OPAL equation of state
\citep{b26} is used. When $\lg T>3.95$ (where $T$ is the temperature), the OPAL opacity
tables \citep{b25, b16} are used; otherwise the opacity tables
from \citet{b1} are used.

As the temperature gradients in the massive stars are essentially independent from what kind of
convection theories is adopted, we apply the Schwarzschild criterion to determine the boundaries
of the convection zones either in the stellar core or in the stellar envelope, and adopt the standard
mixing-length theory to obtain the temperature gradient. The mixing-length parameter $\alpha$ is chosen
to be 1.0 and 0.7, respectively.

When we have obtained the stellar structure model based on above assumptions,
we solve the TCM's equations to obtain the turbulent kinetic energy in the stellar interior.
In the present paper, we choose small values of about $10^{-6}$ for various turbulent correlations
to be the boundary conditions. \textbf{ There are eight parameters in the TCM model we have used: $C_{s}$ is the diffusion parameter for the turbulent kinetic energy; $C_{t1}$ is the diffusion parameter for the turbulent heat flux; $C_{e1}$ is the diffusion parameter of the turbulent temperature fluctuation; $C_{T}$ controls the dissipation rate of the turbulent heat flux; $C_{e}$ controls the dissipation rate of the turbulent temperature fluctuation; $C_{k}$ measures the anisotropic degree of turbulence; $C_{x}$
is the diffusion parameter of turbulent mixing; and $\alpha$ measures the typical length of turbulence. Their suggested values are discussed in detail by \citet{b33}.} In order to see the effects of model parameters on the turbulent properties, we set the values of the turbulent diffusion parameters $C_{s}$, $C_{t1}$, and $C_{e1}$ to
be equal, and use a grid of values to be 0.03, 0.05, and 0.07, respectively. Other parameters of the TCM are chosen as $C_{T}=2.0$, 3.0, and 4.0; $C_{e}=1.00$, 1.25, and 1.50; and $C_{k}=2.0$, 2.5, and 3.0, respectively.

The convective mixing is assumed as usual to be homogeneous in the convective core. But in the overshooting region and in the semi-convection zones, we approximate the convective mixing as a diffusion process, which is described by:
\begin{eqnarray}
&&\frac{\partial X_i}{\partial t}=\frac{\partial}{\partial m}
\left[ \left(4\pi\rho r^2\right)^2 D_t \frac{\partial X_i}{\partial m}\right],
\end{eqnarray}
where $X_i$ is the abundance of species i, and $m$ is the mass interior to the radius $r$. In Eq. (13), the turbulent diffusivity is defined as:
\begin{eqnarray}
&&D_t=C_{x}\frac{\overline{u_{r}^{'}u_{r}^{'}}}{\sqrt{k}}l,
\end{eqnarray}
where $C_x$ is a new model parameter, and $l$ is the mixing-length defined in Eq. (6).
According to \citet{b33}, we choose parameter $C_x$ to be $10^{-9}$, unless otherwise specified elsewhere.

In our stellar models, the metal abundance $Z$ is fixed to be 0.02 and initial hydrogen content $X$ is chosen to be 0.7. \textbf{A stellar model is obtained as follows. The stellar structure model is computed in a usual way, using the MLT in the convective regions. The TCM equations are then solved with a given set of parameters to obtain the turbulent diffusivity. A diffusion equation is then solved to mix materials in the convective core and overshooting region, as well as in the possible semi-convection zones. We do not iterate this process to be fully consistent with the stellar structure equations. Instead, we choose small time steps in the evolution computations to restrict the errors thus introduced.} The stellar models are evolved from the zero-age main-sequence till hydrogen is exhausted at the stellar center.

\section{ Results of $15M_{\odot}$ models }

We have calculated a series of stellar models of $15M_{\odot}$ adopting the Schwarzschild method with the mixing-length parameter $\alpha=1.0$. As shown in Fig.\,1, the radiative temperature gradient is always less than the adiabatic one outside the convective core, which means that the semi-convection does not develop in these stars \citep{b23}. Consequently, the overshooting will be the only significant phenomenon of convection outside the central convective cores in these models.

\begin{figure}
\setlength{\unitlength}{1cm}
\begin{picture}(5,4)
\includegraphics{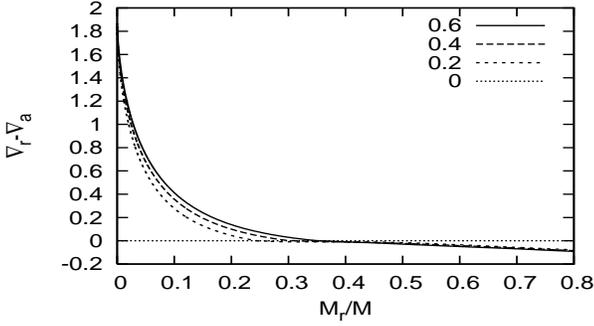}
\end{picture}
\vspace{1cm}
\caption{ Differences of the temperature gradients between the radiative one $\nabla_{r}$ and the adiabatic one $\nabla_{a}$ of  $15M_{\odot}$ models are shown. The turbulent diffusion coefficient $C_{x}$ is $10^{-9}$. The mixing length parameter $\alpha$ is 1.0. Values on the horizontal axis are the mass fractions inside the considered points. The three curves represent three different evolved phases $:$ the full line-the core hydrogen content is 0.6, the dashed line--0.41, the short dashed line--0.2, the dotted line--vertical ordinates' values are 0.}

\end{figure}

\begin{figure}
\setlength{\unitlength}{1cm}
\begin{picture}(5,4)
\includegraphics{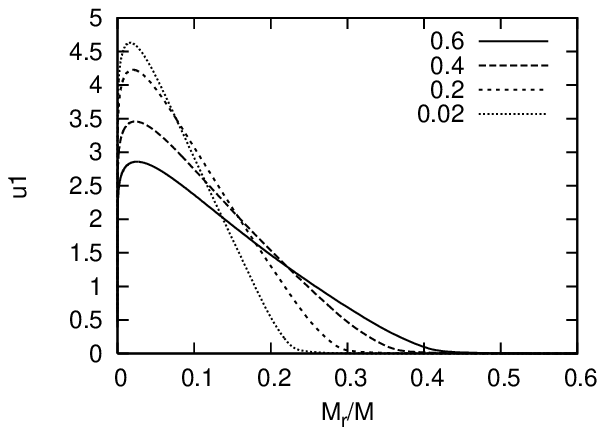}
\put(5,0.){\includegraphics{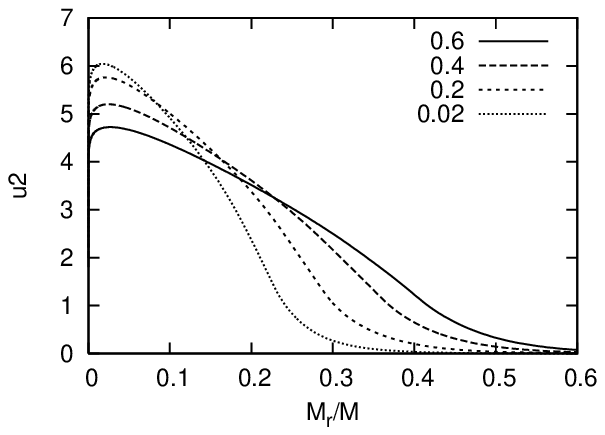}}
\includegraphics{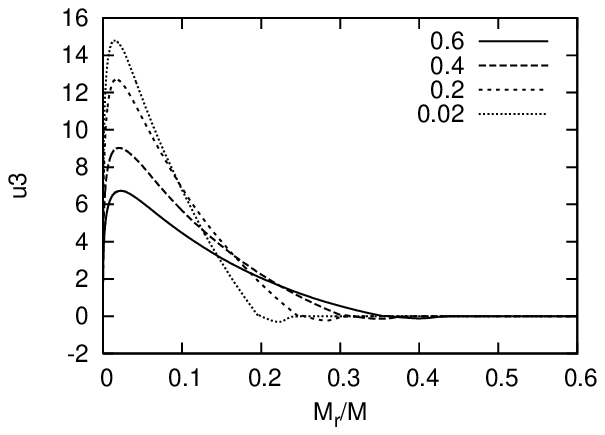}
\put(5,0){\includegraphics{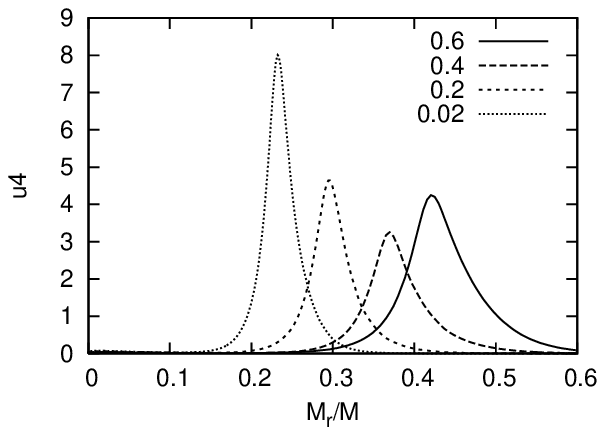}}
\end{picture}
\vspace{15cm}
\caption{ Turbulent properties of $15 M_{\odot}$ models at different evolution stages are shown. Parameters' values are the same with those in Fig. 1. The Schwarzschild method is used to define the convective region and to calculate the turbulent quantities. Values on the horizontal axis are the mass fractions inside the considered points and those on the vertical axis (from top to bottom) are: the radial turbulent dynamic energy $(\overline{u_{r}'u_{r}'})$ (u1/$10^{9}$), the square root of the turbulent dynamic energy $(\sqrt{k})$ (u2/$10^{4}$), the turbulent heat flux $(\overline{u_{r}'T'})$  (u3/$10^{5}$)and the square temperature fluctuation $(\overline{T'T'})$ (u4/$10^{4}$) separately. The four curves represent four different evolved phases: three of them are the same with those in the Fig. 1 and the last one is the phase when the central hydrogen content is 0.02. }
\end{figure}

\subsection{Overshooting from the convective core}

Profiles of the turbulent correlations, computed using the Schwarzschild method,
are shown in Fig.\,2 for four stellar models having the central hydrogen abundance
decreasing equally with time. It can be seen that along with the coverage of the convective core
successively shrinking back, the maximum of the turbulent velocity ($\sqrt{k}$) increases near
the stellar center correspondingly. On the other hand, the turbulent velocity does not drop to zero
at the surface of the convective core, but decays instead continuously in the overshooting region.
This is the direct result of the turbulent diffusion described by the terms on the left hand side
of Eqs.\,(1)-(4). It can be found furthermore, with careful inspections, that the decreasing speed
of the turbulent velocity becomes smaller and smaller when going further and further into the
overshooting region. In contrast, however, the radial kinetic energy of turbulence ($\overline{u_{r}'u_{r}'}$) decays
almost linearly in the overshooting region, which has been used in the diffusion coefficient of the convective mixing there.

It should be noticed in Fig.\,2 that the maximum of the auto-correlation for the
temperature fluctuation ($\overline{T'T'}$) is not inside the convective core, but located
in the overshooting region just outside the convective boundary. This can be easily understood
by inspecting Eq.\,(4). In the convective core, the temperature gradient is almost equal to the
adiabatic one, which results in extremely small turbulent temperature fluctuations. Beyond the
convective core, however, the temperature gradient is considerably smaller than the adiabatic one,
which contributes according to Eq.\,(4) significantly to the generation of the turbulent
temperature fluctuation. From Eqs.\,(1) and (3), such a temperature fluctuation is always a production
factor of the turbulent kinetic energy. As a result, the decay of the turbulent velocity becomes slower
in the overshooting region.

\subsection{Comparisons between the Schwarzschild method and the Ledoux method}

\begin{figure}
\setlength{\unitlength}{1cm}
\begin{picture}(5,4)
\includegraphics{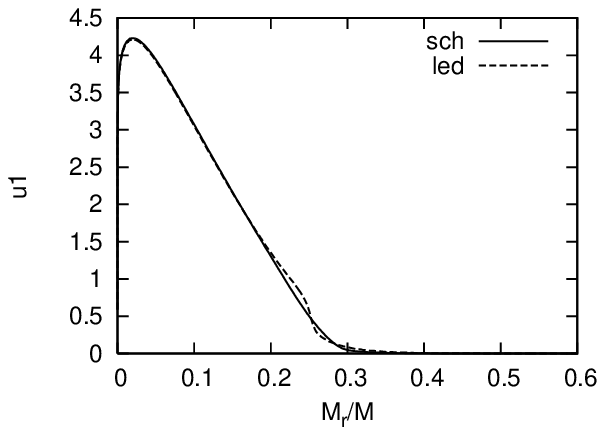}
\put(5,0.){\includegraphics{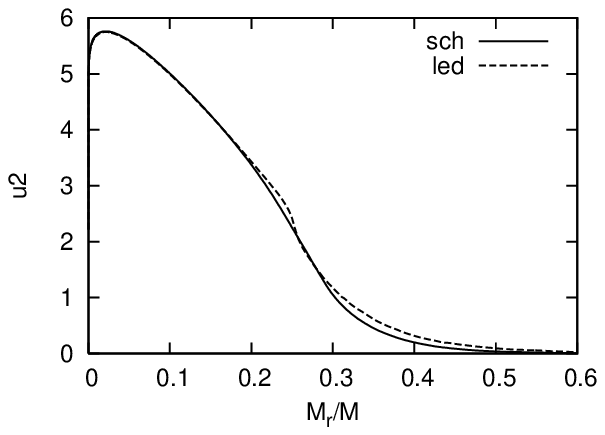}}
\includegraphics{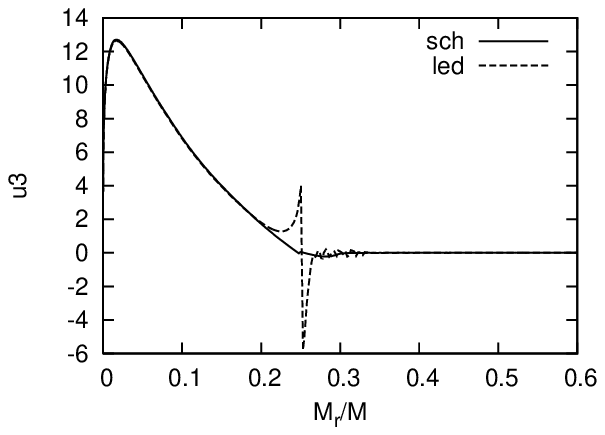}
\put(5,0){\includegraphics{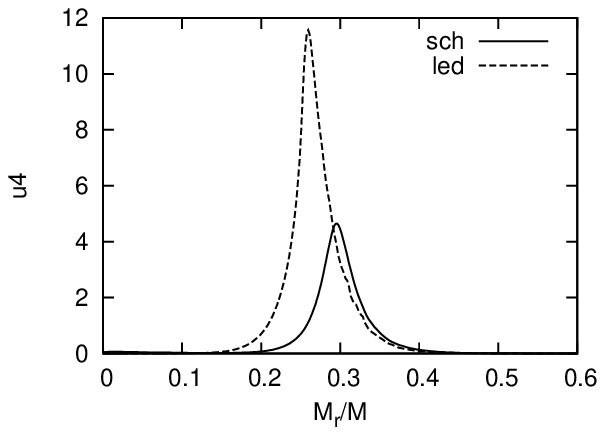}}
\end{picture}
\vspace{15cm}
\caption{ Turbulent properties of $15 M_{\odot}$ models with different convection processing methods are shown.    The evolution phase is when the core hydrogen abundance is 0.2. Parameters' values are the same with those in Fig. 1. The Schwarzschild method and the Ledoux method are used to define the convective region and calculate the turbulent quantities respectively, of which labels are "sch" and "led". Values on the horizontal axis and the vertical axis are the same with those in Fig. 2 except that values' units are different $:$ u4/$10^{4}$--'sch', u4/$10^{6}$--'led'). }
\end{figure}

\begin{figure}
\setlength{\unitlength}{1cm}
\begin{picture}(5,4)
\includegraphics{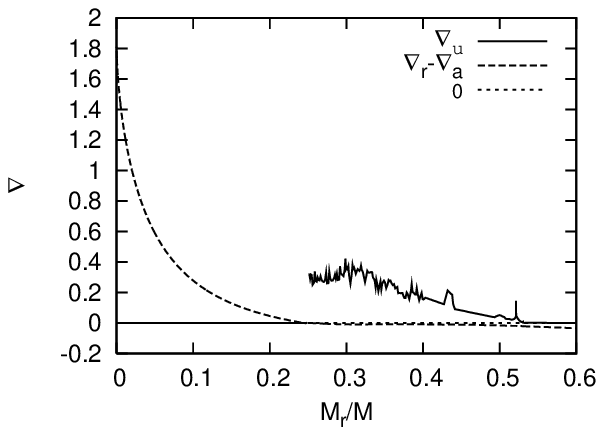}
\put(5,0.){\includegraphics{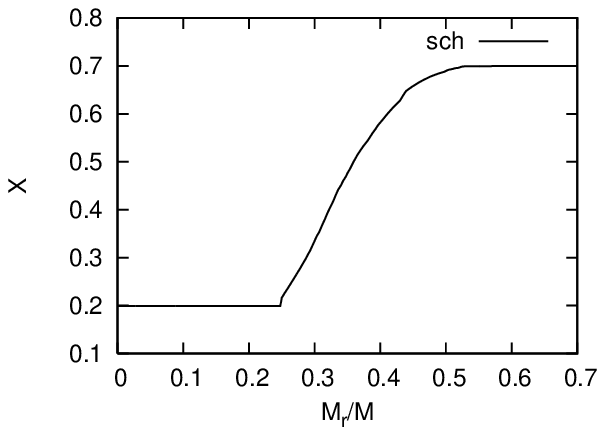}}
\end{picture}
\vspace{5.3cm}
\caption{ Differences of the temperature gradients between the radiative one $\nabla_{r}$ and the adiabatic one $\nabla_{a}$ and the chemical gradient $\nabla_{\mu}$ of $15 M_{\odot}$ models are shown on the upper diagram and the Hydrogen abundance profile is shown on the bottom diagram. Parameters' values are the same with those in Fig. 1. Values on the horizontal axis are the mass fractions inside the considered points. Line of which label is '0' --- the dotted line is a line on which the vertical ordinates' values are 0 }
\end{figure}

\begin{figure}
\setlength{\unitlength}{1cm}
\begin{picture}(5,4)
\includegraphics{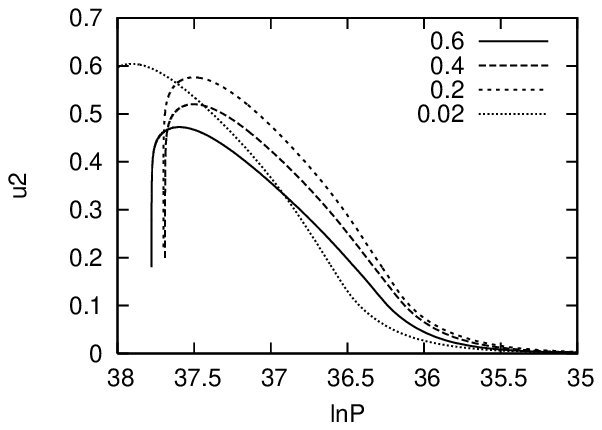}
\put(5,0.){\includegraphics{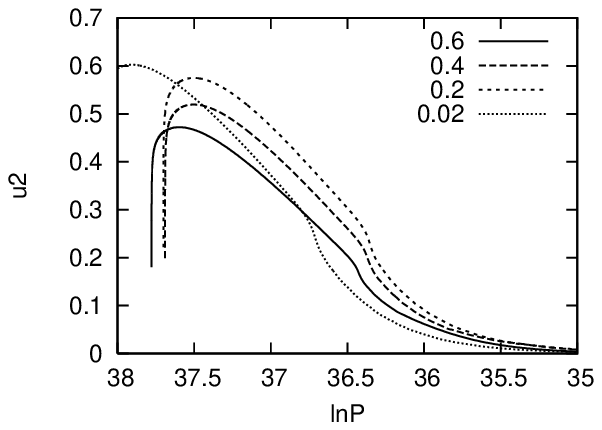}}
\end{picture}
\vspace{5.3cm}
\caption{ The square root of the turbulent dynamic energy $(\sqrt{k})$ (u2/$10^{5}$) of $15 M_{\odot}$ models at different evolution stages are shown. The Schwarzschild method and the Ledoux method are used to define the convective region and calculate the turbulent quantities seperately ( upper--Schwarzschild, bottom--Schwarzschild). Other parameters' values are the same with those in Fig. 1. Values on the horizontal axis are the natural logarithmic pressure. The four curves represent four different evolved phases as explained in Fig. 2.}
\end{figure}

In Fig.\,3, turbulent velocities and auto-correlations of the temperature fluctuation resulted
from the Schwarzschild method and from the Ledoux method are compared. It can be seen that the
two methods lead to exactly the same result in the convective core. But outside the surface of
the convective core, the Ledoux method results in a decay of the turbulent velocity steeper at
first then slower later compared to the result of the Schwarzschild method. On the other hand,
it is striking to notice that the auto-correlation of temperature fluctuation resulted from
the Ledoux method is about two orders of magnitude larger than that resulted from the Schwarzschild method.

These results can be understood by taking the effect of the chemical gradient into account
in the overshooting region. The chemical gradient is shown in Fig.\,4 for the above stellar
model, as well as the difference of the temperature gradients between the radiative one and the
adiabatic one. It can be seen that the chemical gradient is much larger than the difference of the
radiative and adiabatic temperature gradients. According to Eq.\,(3), this fact will directly result
in the depression of the turbulent velocity just beyond the surface of the convective core. On
the other hand, however, the chemical gradient dominantly contributes according to Eq.\,(4) to the
generation of the temperature fluctuation in the overshooting region. As a result, the decay of the
turbulent velocity slows down further in the overshooting region where the temperature fluctuation
reaches its maximum.

The turbulent velocities resulted from the Schwarzschild method and from the Ledoux method are shown
in Fig.\,5 for the stellar models with less and less central hydrogen content. It can be seen that
under the effect of the chemical gradient discussed above, the e-folding distance resulted from
the Ledoux method is about 0.5$H_P$, which is similar to that obtained by the Schwarzschild method.
This indicates that the effect of buoyancy to prevent the motion is largely compensated by the effect
of heating from the convective heat flux against the temperature gradient in the overshooting region.
As the convective mixing in the overshooting region cannot be a fully mixing process, the effective
mixing distance will be significantly shorter than the e-folding distance of the turbulent velocity.

\subsection{Comparisons between results with different mixing-length parameters}

\begin{figure}
\setlength{\unitlength}{1cm}
\begin{picture}(5,4)
\includegraphics{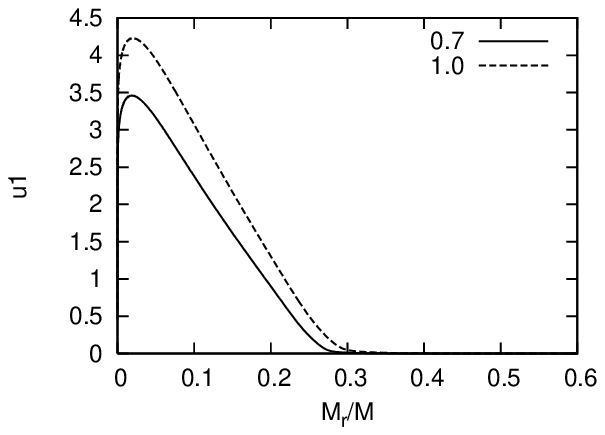}
\includegraphics{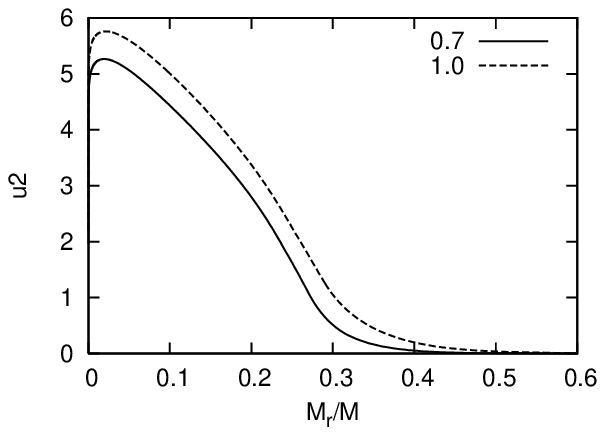}
\includegraphics{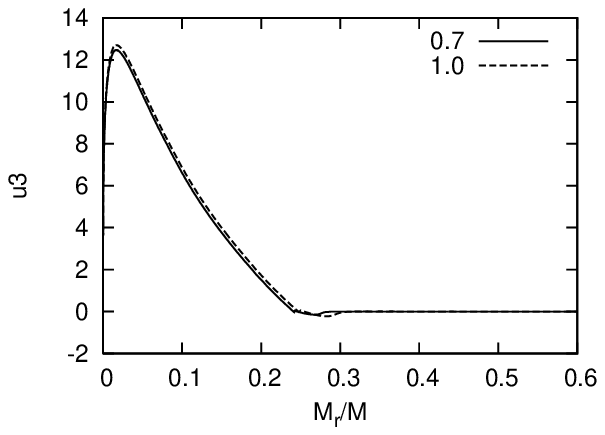}
\includegraphics{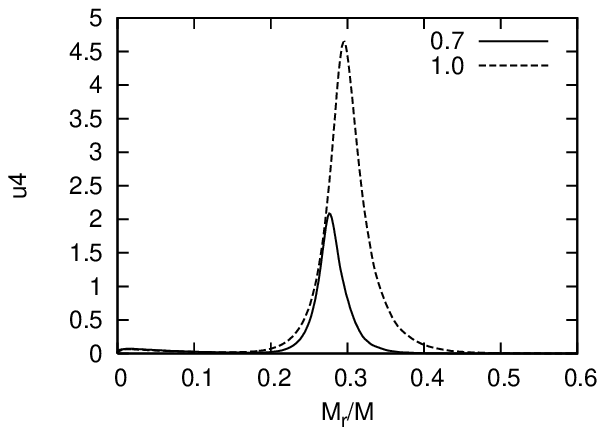}
\end{picture}
\vspace{13.5cm}
\caption{ Turbulent properties of $15 M_{\odot}$ models with different mixing-length parameters ($\alpha$) ('0.7' and '1.0') are shown. The evolution phase is when the core hydrogen abundance is 0.2. The Schwarzschild method is used. Values on the x-axis are the mass fraction inside the considered points. Values on the vertical axis and other parameters' values are the same with those in Fig. 2.}
\end{figure}

It is well known for massive stars that the value of the mixing-length parameter $\alpha$ has little effect on the stellar structure and evolution. However, it may have considerable effect on the turbulence properties in the stellar convective core, as shown in Fig.\,6.

It can be seen that increasing the mixing length parameter $\alpha$ from 0.7 to 1 leads to a considerable increment of the turbulent velocity and radial kinetic energy. From the point of view of fluid dynamics, a larger value of $\alpha$ means that buoyancy can do work on the convective cells for a longer distance, so that the resulted kinetic energy energy of turbulence will be larger. However, the turbulent heat flux ($\overline{u_{r}'T'}$) remains almost the same in the stellar interior, due chiefly to the fact that the temperature gradient is basically adiabatic in the convective core. On the other hand, it is important to notice that there is a significant increment in the auto-correlation of temperature fluctuation in the overshooting region. As a great temperature fluctuation can increase the generation of the turbulent kinetic energy in the overshooting region, a larger mixing length parameter $\alpha$ can therefore slightly increase the overshooting distance beyond the convective core.

\section{Results of $30M_{\odot}$ models}

\subsection{Development of semi-convection and overshooting}

\begin{figure}
\setlength{\unitlength}{1cm}
\begin{picture}(5,4)
\includegraphics{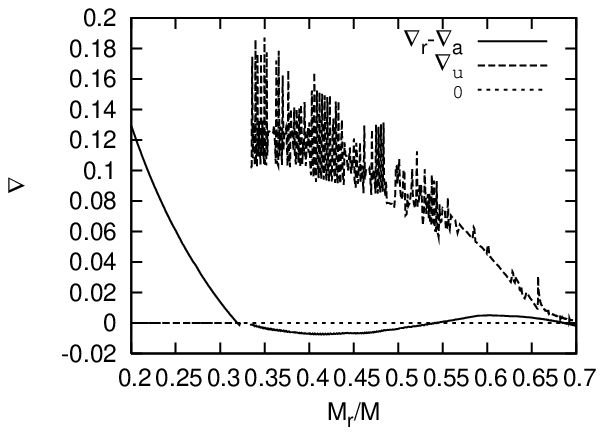}
\put(5,0.){\includegraphics{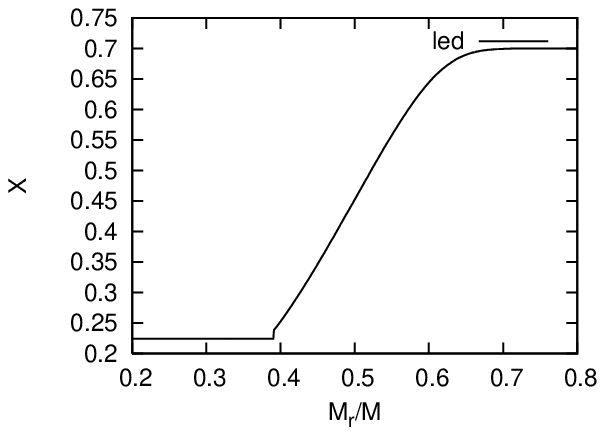}}
\end{picture}
\vspace{5.3cm}
\caption{ Differences of the temperature gradients between the radiative one $\nabla_{r}$ and the adiabatic one $\nabla_{a}$ and the chemical gradient $\nabla_{\mu}$ of $30 M_{\odot}$ models are shown on the upper diagram and the Hydrogen abundance profile is shown on the bottom diagram. The turbulent diffusion coefficient $C_{x}$ is $5\times10^{-9}$. The mixing-length parameter $\alpha$ is 1.0. Values on the horizontal axis and the vertical axis are the same with those in Fig. 4.}
\end{figure}

By use of the Ledoux method, we have computed a series of stellar evolution models of $30M_{\odot}$, with the parameter $C_X=5\times 10^{-9}$. The difference between radiative and adiabatic temperature gradients, as well as the chemical gradient, are shown in Fig.\,7 for a stellar model in late main sequence phase. It can be noticed that the radiative temperature gradient can be greater than the adiabatic one outside the convective core. This is due chiefly to the fact that the opacity, which comes mainly from free-free absorptions and is proportional to the hydrogen abundance, increases outward in the chemical gradient region. As a result, the semi-convection occurs in some part of the chemical gradient region. Overshooting from the convective core is always present. Consequently, these stellar models are a good sample to study interaction between semi-convection and overshooting.

\begin{figure}
\setlength{\unitlength}{1cm}
\begin{picture}(5,4)
\includegraphics{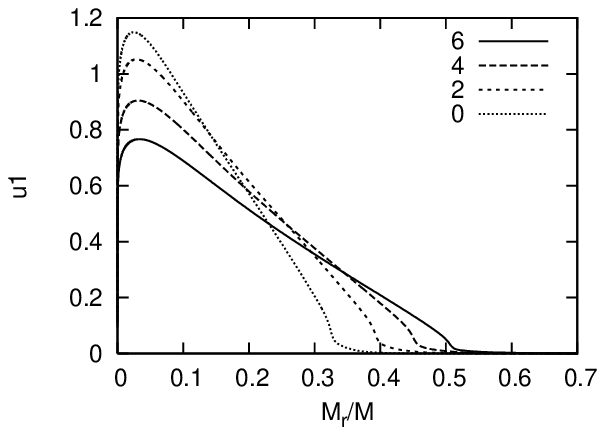}
\put(5,0.){\includegraphics{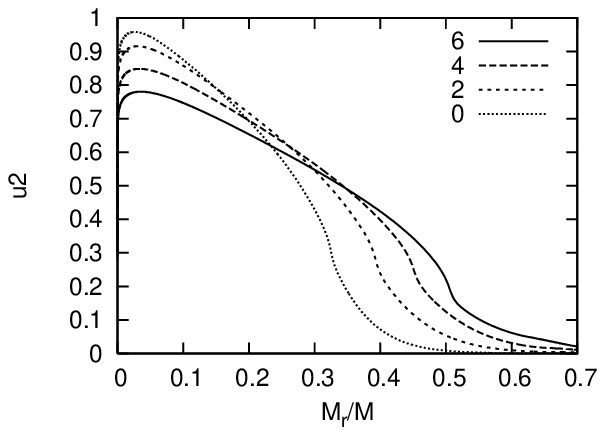}}
\includegraphics{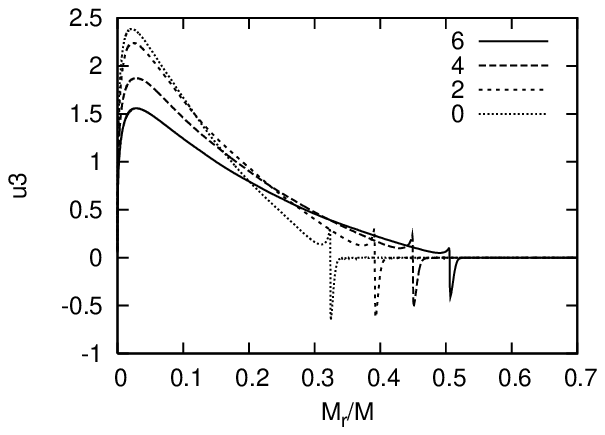}
\put(5,0){\includegraphics{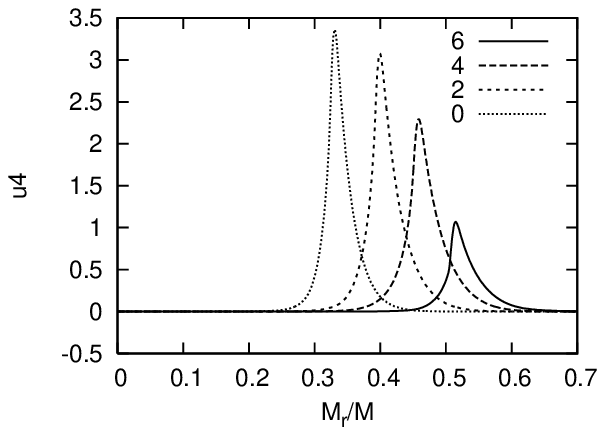}}
\end{picture}
\vspace{13.5cm}
\caption{  Turbulent properties of $30 M_{\odot}$ models at different evolution stages are shown. Parameters' values are the same with those in Fig.7. The Ledoux method is used. Values on the horizontal axis and the vertical axis are the same with those in Fig. 2 except that values' units are different $:$ (u1/$10^{10}$, u2/$10^{5}$, u3/$10^{6}$, u4/$10^{6}$). The four curves represent four different evolved phases $:$ the full line '6' -the core hydrogen content is 0.586, the dashed line '4' --0.4, the short dashed line '2' --0.224, the dotted line '0' --0.03.  }
\end{figure}

\begin{figure}
\setlength{\unitlength}{1cm}
\begin{picture}(5,4)
\includegraphics{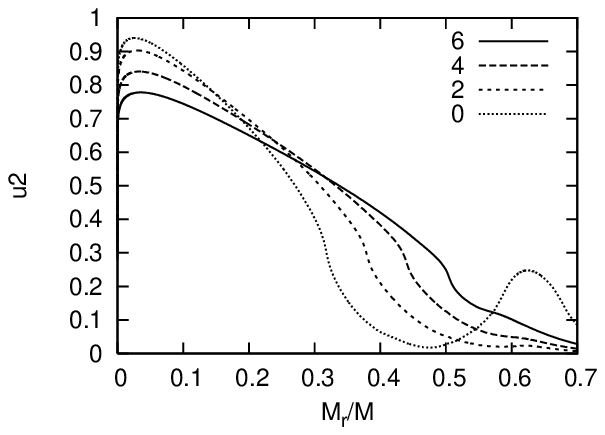}
\end{picture}
\vspace{1cm}
\caption{ The square root of the turbulent dynamic energy $(\sqrt{k})$ (u2/$10^{5}$) of $30 M_{\odot}$ models at different evolution stages are shown. The Ledoux method is used. The element diffusion coefficient $C_{x}$ is $10^{-9}$. The mixing-length parameter $\alpha$ is 1.0. Values on the horizontal axis are the mass fraction inside the considered points and the four curves represent four different evolved phases as explained in Fig. 8.}
\end{figure}

\begin{figure}
\setlength{\unitlength}{1cm}
\begin{picture}(5,4)
\includegraphics{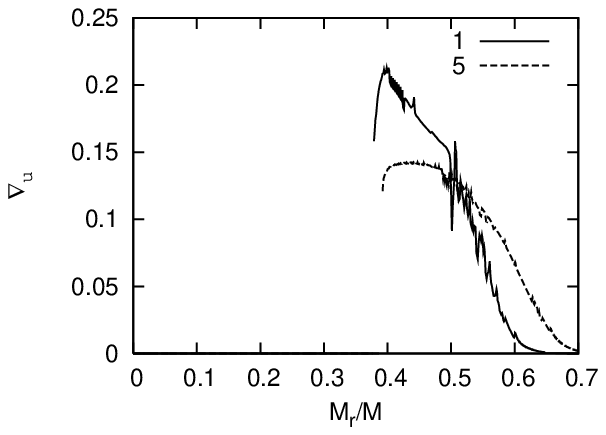}
\put(5,0.){\includegraphics{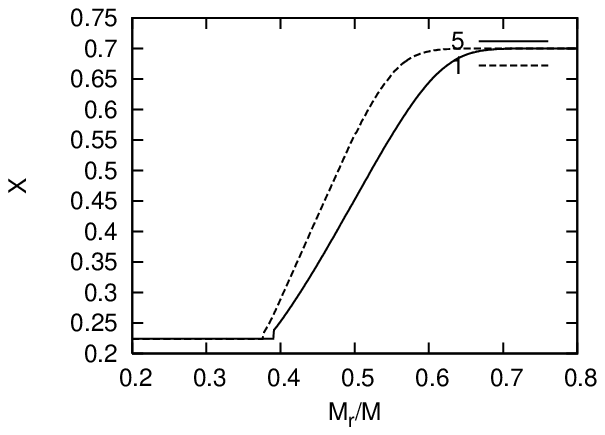}}
\end{picture}
\vspace{5.3cm}
\caption{ The chemical gradients $\nabla_{\mu}$ of $30 M_{\odot}$ models with different turbulent diffusion coefficient ('1'---$C_{x}=1\times10^{-9}$, '5'---$C_{x}=5\times10^{-9}$) are shown on the upper diagram and the Hydrogen abundance profile is shown on the bottom diagram. The mixing-length parameter $\alpha$ is 1.0. The evolution phase is when the core hydrogen abundance is 0.2. Values on the x-axis are the mass fraction inside the considered points.}
\end{figure}

\begin{figure}
\setlength{\unitlength}{1cm}
\begin{picture}(5,4)
\includegraphics{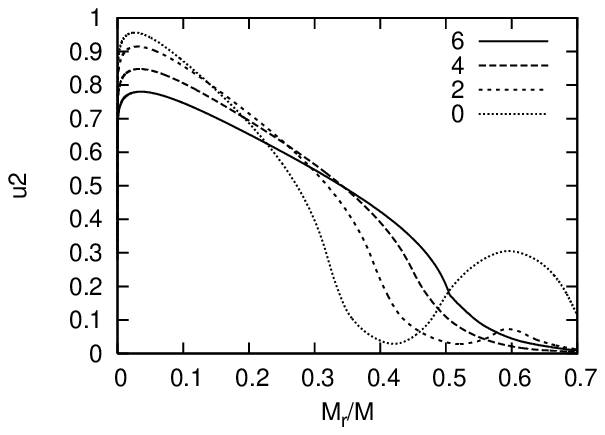}
\end{picture}
\vspace{1cm}
\caption{ The square root of the turbulent dynamic energy $(\sqrt{k})$ (u2/$10^{5}$) of $30 M_{\odot}$ models at different evolution stages are shown. The Schwarzschild method is used. The turbulent diffusion coefficient $C_{x}$ is $5\times10^{-9}$. The mixing length parameter $\alpha$ is 1.0. Values on the horizontal axis are the mass fraction inside the considered points. The four curves represent four different evolved phases as explained in Fig. 8.}
\end{figure}

Profiles of turbulent correlations are shown in Fig.\,8 for some stellar models with nearly equally spaced central hydrogen abundance. It can be seen that the overall properties of convection are similar to the case of the stellar models with $15M_{\odot}$. The convective core shrinks continuously, along with successively increasing turbulent kinetic energy near the stellar center. The overshooting from the convective core is significant, its extension being effectively influenced by the chemical gradient. In particular, the semi-convection indeed develops near the surface of the chemical gradient region for the model with the lowest central hydrogen content, but it actually results in almost no effect on the turbulent velocity for this stellar model.

In order to see how the efficiency due to partial mixing of convection can play a significant role in the chemical gradient region, we have computed one more series of stellar evolution models with $C_X=10^{-9}$. It can be seen in Fig.\,9 that decreasing $C_X$ can significantly magnify the semi-convection, sometimes even driving it into a full convection in the most upper part of the chemical gradient region during late stage of the main sequence. According to Eq.\,(14), the turbulent diffusivity will be smaller if $C_X$ decreases, leading to a weaker mixing outside the convective core. As a result, the chemical gradients will be larger for these stellar models as shown in Fig.\,10, for it is mainly determined by the overshooting beyond the convective core in the preceding stage of the main sequence evolution. Higher hydrogen abundance in the outer part of the chemical gradient region results therefore in larger opacity and larger radiative temperature gradient, which is responsible for driving stronger semi-convection as well as full convection observed in late stage of the main sequence.

In contrast, however, full convection always develops in the chemical gradient regions for stellar models adopting the Schwarzschild method in late main sequence stage, which results in an intermediate convection zone as seen in Fig.\,11 (see also \citet{b23}). Due to exclusion of the chemical gradient term, the semi-convection is not considered in the Schwarzschild method. It can therefore be noticed in Fig.\,11 that the upward overshooting from the convective core penetrates extensively and finally meets the downward overshooting from the intermediate convection zone, resulting in a much stronger mixing in almost all the chemical gradient region.

\subsection{Comparisons between the Schwarzschild method and the Ledoux method}

Turbulent properties are compared for stellar models based on the Ledoux method and on the Schwarzschild method. It can be seen in Fig.\,12 that the Ledoux method gives a slower decay of the turbulent velocity just above the boundary of the convective core than the Schwarzschild method does, due to the similar reason as for the $15M_{\odot}$ model. It can be noticed that the auto-correlation of temperature fluctuation according to the Ledoux method is again much larger than that according to the Schwarzschild method. When going into the semi-convection region, however, the turbulent velocity continues to decay according to the Ledoux method, but with an even slower speed due to the radiative temperature gradient now being larger than the adiabatic one. On the other hand, the turbulent velocity shows instead the second maximum in the same region according to the Schwarzschild method, for the stratification is now unstable against convection if the chemical gradient is omitted.

\begin{figure}
\setlength{\unitlength}{1cm}
\begin{picture}(5,4)
\includegraphics{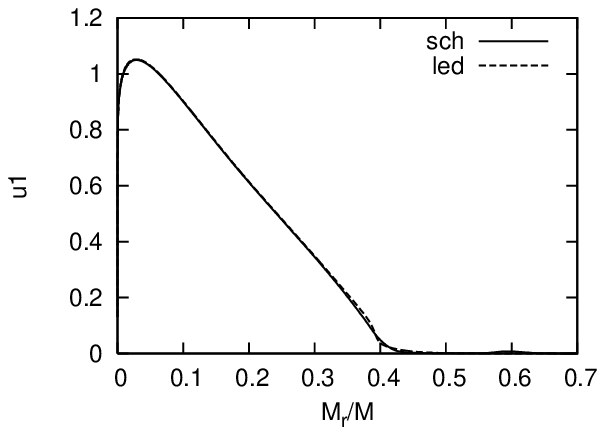}
\put(5,0.){\includegraphics{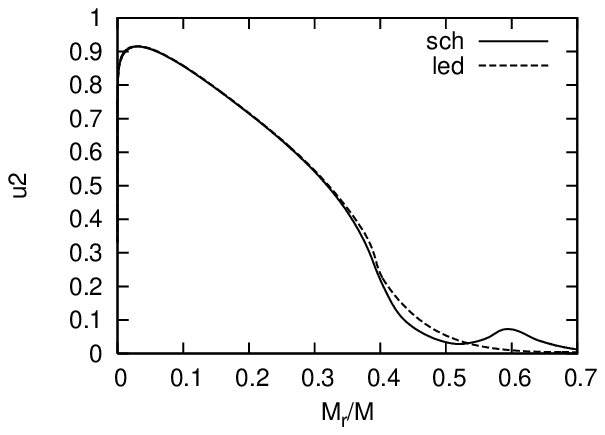}}
\includegraphics{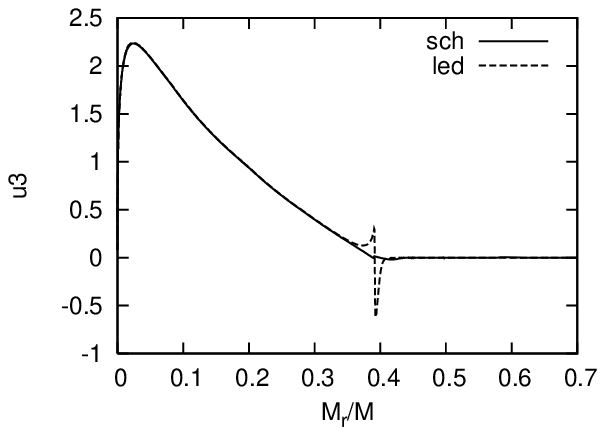}
\put(5,0){\includegraphics{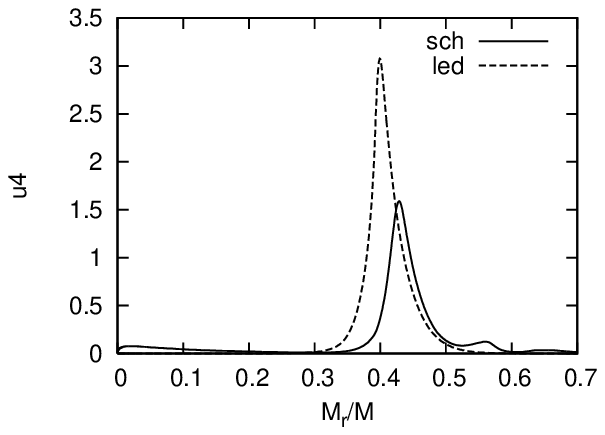}}
\end{picture}
\vspace{15cm}
\caption{ Turbulent properties of $30 M_{\odot}$ models with different convection processing methods are shown.  The evolution phase is when the core hydrogen abundance is 0.2. Parameters' values are the same with those in Fig. 7. The Schwarzschild method and the Ledoux method is used to define the convective region and calculate the turbulent quantities seperately, of which labels are "sch" and "led". Values on the horizontal axis and the vertical axis are the same with those in Fig. 3 except that values' units are different $:$ (u1/$10^{10}$, u2/$10^{5}$, u3/$10^{6}$, u4/$10^{4}$--'sch', u4/$10^{6}$--'led').}
\end{figure}

\subsection{Comparisons between results with different mixing-length parameters}

\begin{figure}
\setlength{\unitlength}{1cm}
\begin{picture}(5,4)
\includegraphics{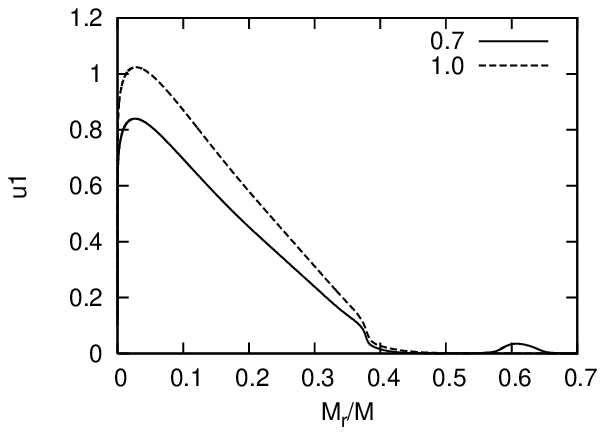}
\put(5,0.){\includegraphics{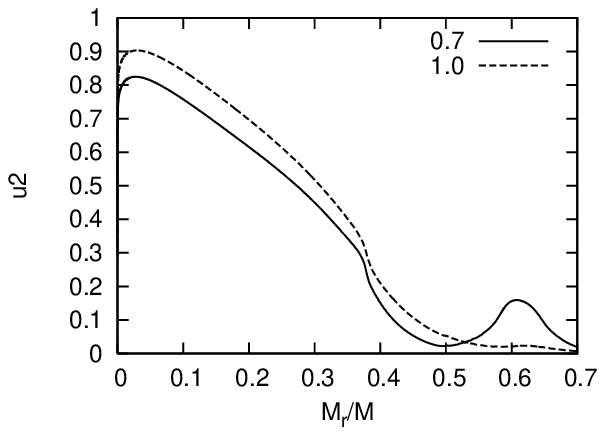}}
\includegraphics{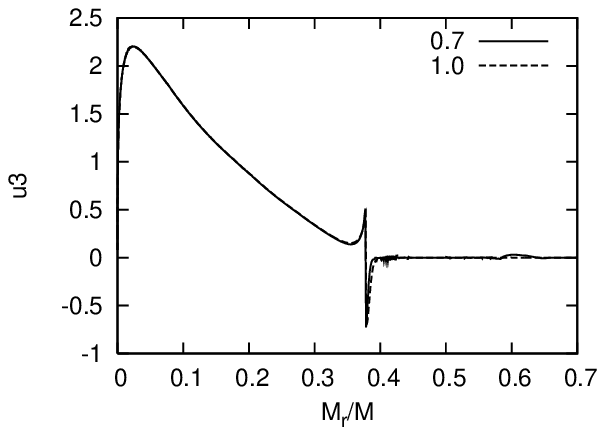}
\put(5,0){\includegraphics{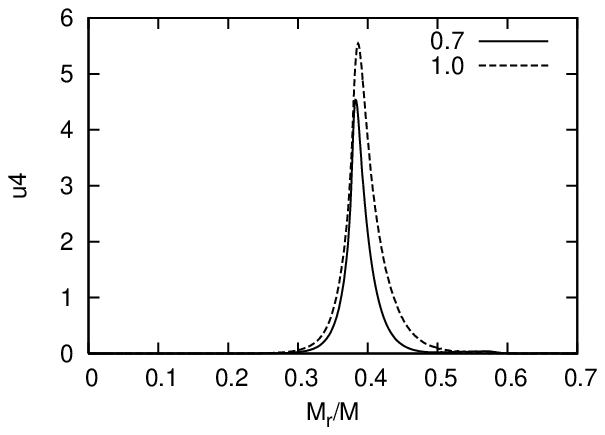}}
\end{picture}
\vspace{15cm}
\caption{ Turbulent properties of $30 M_{\odot}$ models with different mixing-length parameters ($\alpha$) ('0.7' and '1.0') are shown. The evolution phase is when the core hydrogen abundance is 0.2. The Ledoux method is used. The turbulent diffusion coefficient $C_{x}$ is $10^{-9}$. Values on the horizontal axis and the vertical axis are the same with those in Fig. 6 except that values' units are different $:$ (u1/$10^{10}$, u2/$10^{5}$, u3/$10^{6}$, u4/$10^{6}$).}
\end{figure}

Profiles of different turbulent correlations are shown in Fig.\,13 for stellar models with different mixing-length parameter $\alpha$. It can be noticed that increasing $\alpha$ from 0.7 to 1.0 results in similar increments of the turbulent kinetic energy in the convective core, just as for the model of $15M_{\odot}$. However, the turbulent heat flux can hardly be influenced by different choices of $\alpha$, due to the actual temperature gradient being almost equal to the adiabatic one. Just beyond the convective core, the temperature fluctuation is considerably enhanced in the overshooting region by a larger value of $\alpha$, which prolongs noticeably the overshooting distance of the turbulent velocity.

When going further into the semi-convection region, a larger value of $\alpha$ effectively suppresses convective motions there, resulting in almost no intermediate convection zone to appear. This result can be understood by considering the following two aspects. On one hand, increasing $\alpha$ will lead according to Eqs.\,(5) and (6) to a smaller dissipation rate of the turbulent kinetic energy. This effect slows down the decaying rate of the turbulent velocity into the overshooting region and helps the convective motions to penetrate all the way into the semi-convection region. On the other hand, increasing $\alpha$ produces also a more efficient mixing according to Eq.\,(14) in the chemical gradient region, which will result in a smaller chemical gradient and effectively suppress the transition of the semi-convection into a full convection in the chemical gradient region.

\subsection{Turbulent properties among models with different turbulent parameters}

Turbulent properties in the chemical gradient region are sensitively determined by the parameters in the turbulence model we have adopted. From Eqs.\,(1)-(4), we can notice that there are six model parameters: $C_s$, $C_{t1}$, and $C_{e1}$ are diffusion parameters, $C_t$ and $C_e$ are dissipation parameters, and $C_k$ is the coefficient for turbulence anisotropy.

\begin{figure}
\setlength{\unitlength}{1cm}
\begin{picture}(5,4)
\includegraphics{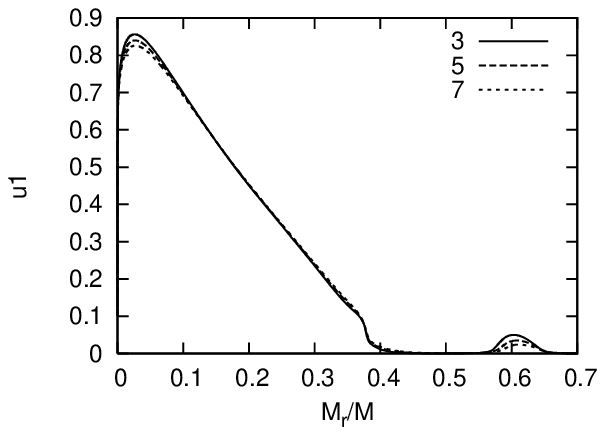}
\put(5,0.){\includegraphics{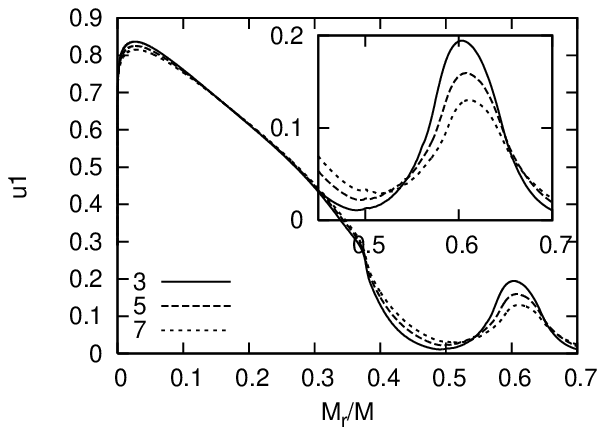}}
\includegraphics{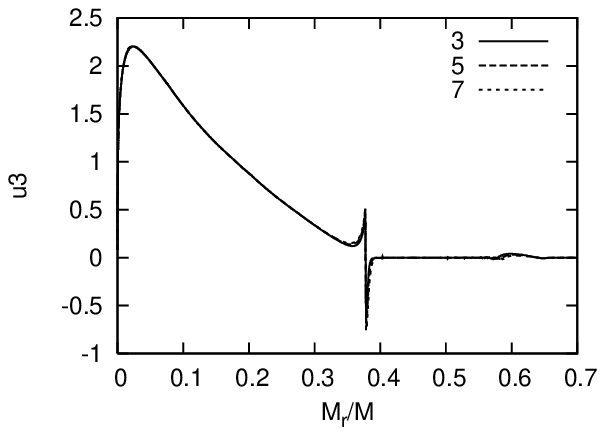}
\put(5,0){\includegraphics{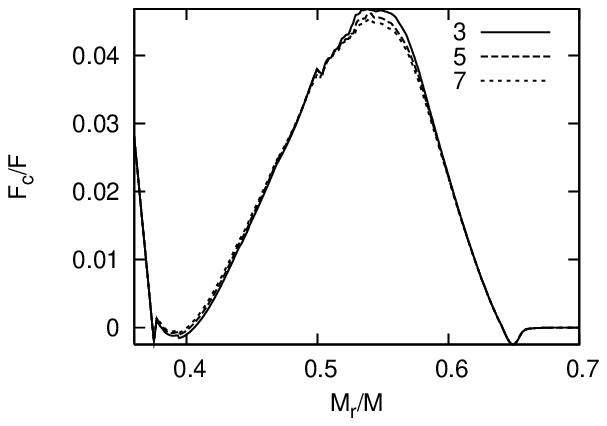}}
\end{picture}
\vspace{15cm}
\caption{ Turbulent properties of $30 M_{\odot}$ models with different turbulent diffusion parameters $C_{s},C_{t1},C_{e1}$ ('3'--0.03, '5'--0.05, '7'--0.07) are shown. The evolution phase is when the core hydrogen abundance is 0.2. The Ledoux method is used. The mixing-length parameter $\alpha$ is 0.7, the turbulent diffusion coefficient $C_{x}$ is $10^{-9}$. Values on the horizontal axis and the vertical axis are the same with those in Fig. 13 except that the vertical ordinates of the bottom diagram is the fraction of convection transported energy to the total energy $(F_{c}/F)$.}
\end{figure}

It can be seen in Fig.\,14, that increasing the turbulent diffusion parameters $C_{s}$, $C_{t1}$, and $C_{e1}$ the curves of turbulent velocity become flatter and flatter, the peak values becoming smaller while the bottom values becoming larger. This property indicates that larger turbulent diffusion parameters lead to more significant overshooting of the turbulent kinetic energy.

\begin{figure}
\setlength{\unitlength}{1cm}
\begin{picture}(5,4)
\includegraphics{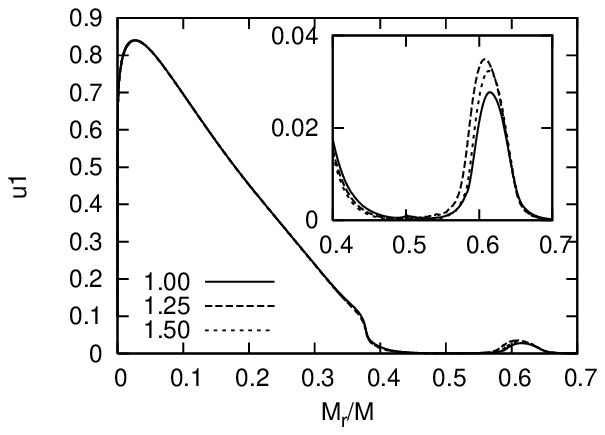}
\put(5,0.){\includegraphics{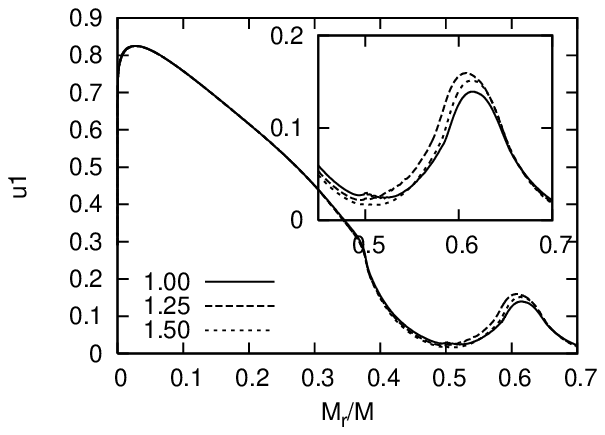}}
\includegraphics{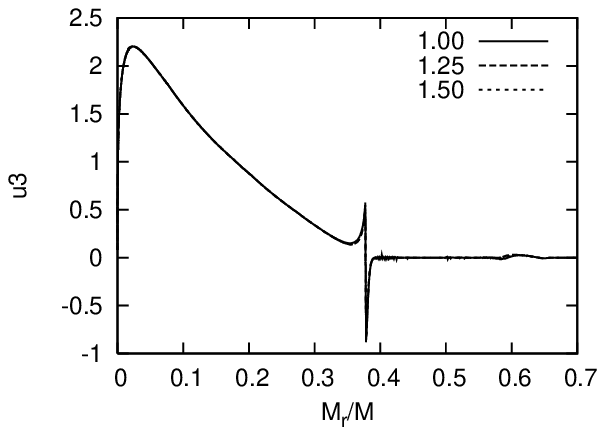}
\put(5,0){\includegraphics{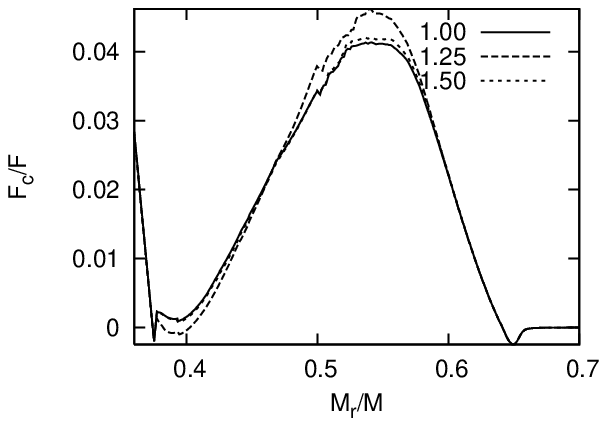}}
\end{picture}
\vspace{15cm}
\caption{ Turbulent properties of $30 M_{\odot}$ models with different turbulent dissipation coefficient $C_{e}$ ('1.00', '1.25', '1.50') are shown. Other values are the same with those in Fig. 14. }
\end{figure}

\begin{figure}
\setlength{\unitlength}{1cm}
\begin{picture}(5,4)
\includegraphics{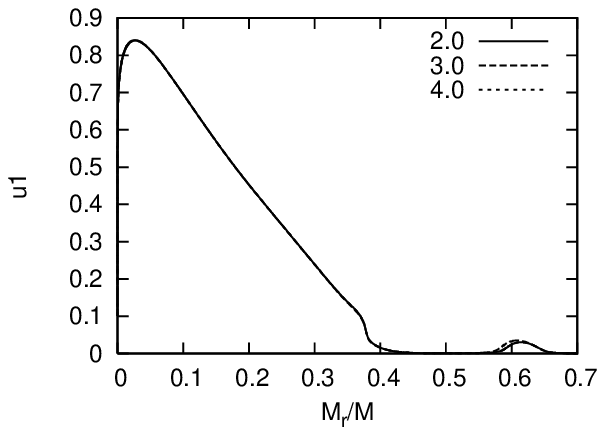}
\put(5,0.){\includegraphics{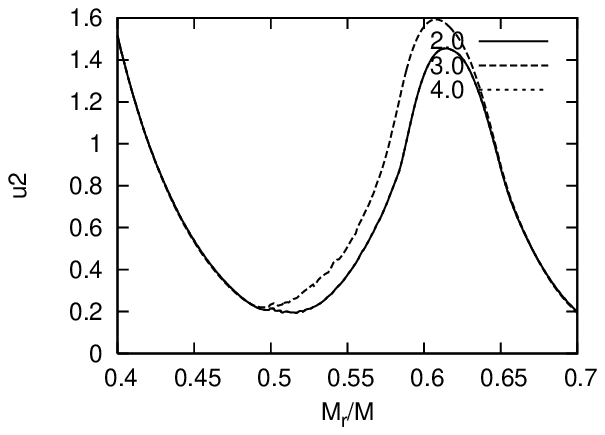}}
\includegraphics{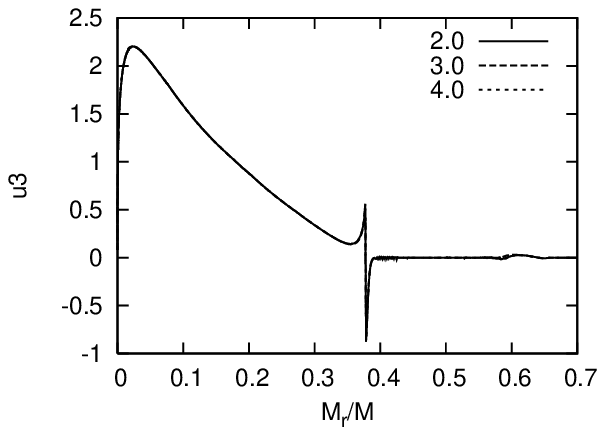}
\put(5,0){\includegraphics{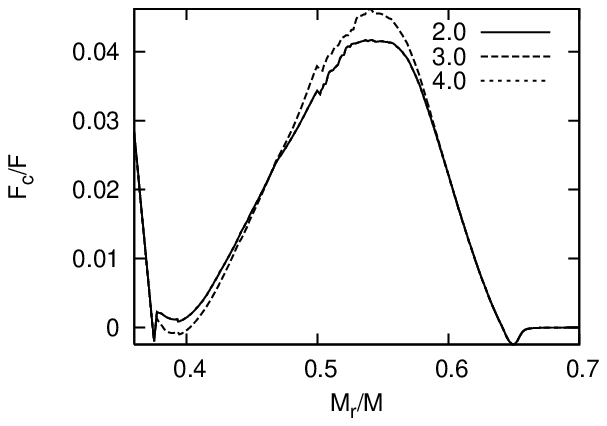}}
\end{picture}
\vspace{15cm}
\caption{ Turbulent properties of $30 M_{\odot}$ models with different turbulent dissipation coefficient $C_{t}$ ('2.0', '3.0', '4.0') are shown. Other values are the same with those in Fig. 14 except that values' units of the top right picture is $10^{4}$. }
\end{figure}

It can be seen in Fig.\,15 that increasing the turbulent dissipation coefficient $C_{e}$ results in little changes for the turbulent velocity $(\sqrt{k})$ in the convective core. In the chemical gradient region, however, the turbulent velocity becomes smaller in the overshooting region just beyond the boundary of the convective core, and then larger in the semi-convection region as well as fully convective intermediate zone. From Eq.\,(4), a larger value of $C_{e}$ is equivalent to a larger dissipation rate $\varepsilon$, which will lead to a weaker overshooting in the turbulent velocity. As a result, mixing previously in the chemical gradient region is weaker, leaving a larger chemical gradient to benefit full convection developing later in the chemical gradient region.

In Fig.\,16, it can be seen that increasing the turbulent dissipation coefficients $C_{t}$ results in a similar effect as increasing $C_e$. The overshooting beyond the convective core is suppressed, while the semi-convection in the outer part of the chemical gradient region is enhanced.

\begin{figure}
\setlength{\unitlength}{1cm}
\begin{picture}(5,4)
\includegraphics{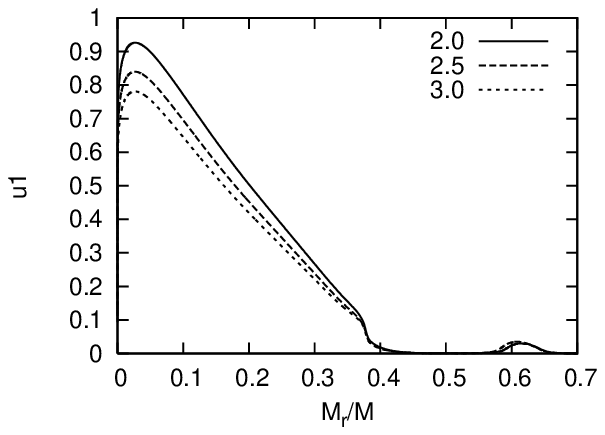}
\put(5,0.){\includegraphics{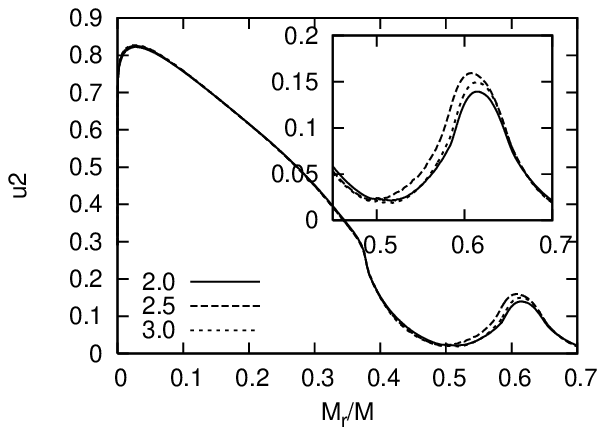}}
\includegraphics{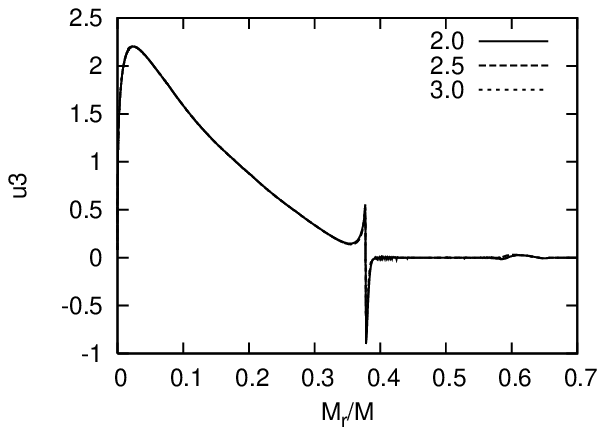}
\put(5,0){\includegraphics{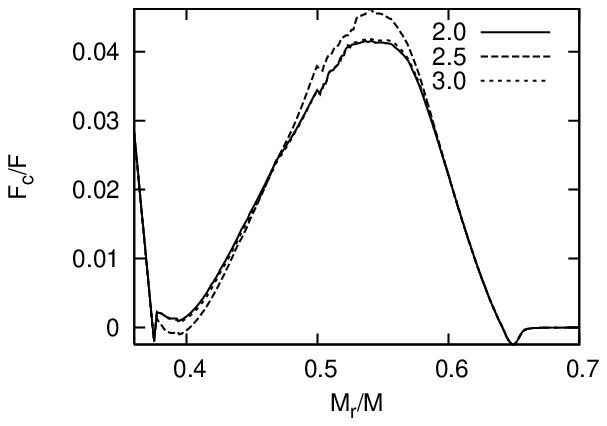}}
\end{picture}
\vspace{15cm}
\caption{ Turbulent properties of $30 M_{\odot}$ models with different turbulent redistribution coefficient $C_{k}$ ('2.0', '2.5', '3.0') are shown. Other values are the same with those in Fig. 14. }
\end{figure}

It can be seen in Fig.\,17 that in the convective core, increasing the redistribution coefficient of  turbulence $C_{k}$ reduces the radial kinetic energy $(\overline{u_{r}'u_{r}'})$, while it does not affect the turbulent velocity. This is because that $C_{k}$ determines how the kinetic energy energy is distributed among different direction of motions. In the chemical gradient region, however, increasing $C_{k}$ leads to a considerable increment of the turbulent velocity. This effect can be understood by considering Eq.\,(3). In the chemical gradient region, the buoyancy prevents the convective motions and results in a smaller kinetic energy in the radial direction than in the other directions. A larger value of $C_{k}$ tends to make the turbulent motions more isotropic, so that the radial kinetic energy will be accordingly larger. From Eq.\,(3), a larger $(\overline{u_{r}'u_{r}'})$ implies a larger $\varepsilon$, leading to a smaller turbulent velocity in the overshooting region. As a result, the mixing is weaker in the overshooting region to give a larger chemical gradient in the previous evolution stage, which will be responsible for a stronger semi-convection in late stage of the main sequence.

In Figs.\,14-17, we can see that the temperature-velocity correlation of turbulence $(\overline{u_{r}'T'})$ is hardly influenced by different choices of the parameters of our turbulence model, and the resulted convective heat flux is essentially unaffected.

\section{Summary and Discussions}

Referring to the Schwarzschild criterion and Ledoux criterion, there are the Schwarzschild method and Ledoux method which we use to confine the convective boundary and calculate the turbulent properties.
Models in this paper are different from \citet{b12} and  \citet{b13} that the turbulent velocity we used to calculated the turbulent diffusion  are from TCM and theirs are from phenomenological models. And the entropy gradient $\frac{\partial \overline{s}}{\partial r}=-\frac{c_{p}}{H_P}(\nabla-\nabla_{\rm ad}-\nabla_{\mu})$ which takes the chemical gradient into account is different from \citet{b20} and \citet{b33}.

In the present paper, turbulent quantities for several series of models for stellar with $30M_{\odot}$ and $15M_{\odot}$ are calculated.
From our calculations, semi-convection appears in models for stellar with $30M_{\odot}$ but doesn't appear in models for stellar with $15M_{\odot}$.	
With the central hydrogen abundance decreasing, the coverage of the convective core successively shrinks back, and the maximum of the turbulent velocity increase near the stellar center while the turbulent velocity
decays continuously in the overshooting region. Owing to the production effect of the turbulent temperature fluctuation, the decay
of the turbulent velocity becomes slower when going further and further into the overshooting region.
For stellar models of $30M_{\odot}$, semi-convection occurs in some part of the chemical gradient region due to
the opacity which comes mainly from free-free absorptions increasing outward in the chemical gradient region.

We analyzed the influences of different parameters on the turbulent quantities. In general, we can summarize as following:

1) In this paper, it can be found that the semi-convection occurring in late stage of the main sequence
is closely related to the overshooting beyond the convective core in early stage of the main sequence.
A stronger overshooting in the previous evolution results in a stronger mixing in the chemical gradient
region and a smaller chemical gradient, leaving a lower hydrogen abundance and a smaller radiative temperature
gradient in the outer part of the chemical gradient region, which finally leads to a weaker semi-convection in
late stage of the main sequence.

2)	The Schwarzschild method adopted to define the convective region and to calculate the turbulent quantities
 is more conductive to occurrence of semi-convection than the Ledoux method adopted. The overshooting distance
 is further and the auto-correlation of temperature fluctuation is depressed when adopting the Ledoux method.
 From models for stellar with $15M_{\odot}$, the effective mixing distance in the overshooting region is significantly
 shorter than $0.5H_{p}$ adopting either method.

3)	Increasing the mixing length parameter $\alpha$ leads to increasing the turbulent dynamic energy in
the convective core and prolonging the overshooting distance but decreasing the turbulent dynamic energy
in the semi-convective zone. This results can be understood by considering two aspects: on one hand,
increasing $\alpha$ leads to a smaller dissipation rate of turbulent dynamic energy which slows down the decaying rate
of the turbulent velocity in the overshooting region; on the other hand, increasing $\alpha$ produces a more efficient
mixing in the chemical gradient region which results in a smaller chemical gradient and suppresses the transition of
the semi-convection into a full convection.

4)	Increasing the turbulent diffusion parameters ($C_{s}, C_{t1}, C_{e1}$) devote to flatten
the curves of the turbulent dynamic energy and lead to more significant overshooting of turbulent dynamic
energy. When increasing the turbulent dissipation coefficient $C_{e}$ or $C_{t}$ , the turbulent dynamic
 energy is smaller in the overshooting region but larger in the semi-convective region as well as in the
 intermediate convective zones. When increasing the turbulent redistribution coefficient $C_{k}$, the
 radial turbulent dynamic energy is smaller but the turbulent dynamic energy doesn't change in the
 convective core, while in the chemical gradient region, the turbulent dynamic energy is smaller in
 the overshooting region and larger in the semi-convective region as well as in the intermediate
  convective zones.  The temperature-velocity correlation of turbulence $(\overline{u_{r}'T'})$ is
  hardly influenced by different choices of the parameters of our turbulence model, and the resulted
  convective heat flux is essentially unaffected.

\section*{Acknowledgments}
 Very appreciate my fellow apprentices' (Zhang Qian sheng, Lai xiang jun, Su jie, et al) sharing experiences and help in working and life.
This work is co-sponsored by the NSFC of China (Grant Nos. 11333006 and 10973035),
and by the Chinese Academy of Sciences (Grant No. KJCX2-YW-T24).

\label{lastpage}

\end{document}